# Observation and control of maximal Chern numbers in a chiral topological semimetal

*Niels B.M. Schröter[1,*], Samuel Stolz[2,3], Kaustuv Manna[4], Fernando de Juan[5,6], Maia G. Vergniory[5,6], Jonas A. Krieger[1,7,8], Ding Pei[9], Thorsten Schmitt[1], Pavel Dudin[10,***], Timur K. Kim[10], Cephise Cacho[10], Barry Bradlyn[11], Horst Borrmann[4], Marcus Schmidt[4], Roland Widmer[2], Vladimir N. Strocov[1], and Claudia Felser[4,**].*

[1]*Swiss Light Source, Paul Scherrer Institute, CH-5232 Villigen PSI, Switzerland*
[2]*EMPA, Swiss Federal Laboratories for Materials Science and Technology, 8600 Dübendorf, Switzerland*
[3]*Institute of Condensed Matter Physics, Station 3, EPFL, 1015 Lausanne, Switzerland*
[4]*Max Planck Institute for Chemical Physics of Solids, Dresden, D-01187, Germany*
[5]*Donostia International Physics Center, 20018 Donostia-San Sebastian, Spain*
[6]*IKERBASQUE, Basque Foundation for Science, Maria Diaz de Haro 3, 48013 Bilbao, Spain*
[7]*Laboratory for Muon Spin Spectroscopy, Paul Scherrer Institute, CH-5232 Villigen PSI, Switzerland*
[8]*Laboratorium für Festkörperphysik, ETH Zurich, CH-8093 Zurich, Switzerland*
[9]*Clarendon Laboratory, Department of Physics, University of Oxford, Oxford OX1 3PU, United Kingdom*
[10]*Diamond Light Source, Didcot, OX110DE, United Kingdom*
[11]*Department of Physics and Institute for Condensed Matter Theory, University of Illinois at Urbana-Champaign, Urbana, IL, 61801-3080, USA*

Correspondence to: [*]*niels.schroeter@psi.ch*

[**]*Claudia.Felser@cpfs.mpg.de*

[***] *Current affiliation:* Synchrotron Soleil, 91192 Gif-sur-Yvette, France

**Topological semimetals feature protected nodal band degeneracies characterized by a topological invariant known as the Chern number (C). Nodal band crossings with linear dispersion are expected to have at most |C|=4, which sets an upper limit to the magnitude of many topological phenomena in these materials. Here we show that the chiral crystal PdGa displays multifold band crossings, which are connected by exactly four surface Fermi-arcs, thus proving that they carry the maximal Chern number magnitude of 4. By comparing two enantiomers, we observe a reversal of their Fermi-arc velocities, which demonstrates that the handedness of chiral crystals can be used to control the sign of their Chern numbers.**

Topological invariants are mathematical objects that can be used to classify Hamiltonians, and have found widespread applications in physics, chemistry, and materials science. One of the best known topological invariants in condensed matter physics is the Chern number, which can be defined as the flux of Berry curvature through a closed two-dimensional surface. If this surface is taken to be the whole Brillouin zone, the Chern number classifies insulators in two dimensions, as first used in the context of the quantum Hall effect by Thouless and co-workers (*1*, *2*) in the 1980s. More recently, Chern numbers have also been used to classify topological nodal semimetals (*3*), where point-like energy degeneracies in their bulk electronic structure act as sources and sinks of quantized Berry flux through any local isoenergy surface enclosing the node. For the simplest case of a linear touching of two bands, which can occur in any non-centrosymmetric or magnetic material and is called a Weyl point, the magnitude of the Chern number C is limited to |C|=1. However, there is nothing preventing more complicated nodal crossings from having larger Chern numbers; this has important consequences, as the magnitude of many of the exotic phenomena predicted for topological semimetals is directly proportional to their Chern number. Examples include the number of topological Fermi-arc surface states (*3, 4*), the number of chiral Landau levels influencing magnetotransport phenomena related to the chiral anomaly (*5, 6*), the magnitude of the quantized rate of photocurrents in the quantized circular photogalvanic effect (*7–9*), and many more (*10–12*). Owing to the importance of the Chern number magnitude for these phenomena, it is natural to ask whether there is an upper limit for this topological invariant, and whether there are real materials in which this limit can be reached.

It has recently been predicted that in chiral crystals, which possess neither mirror nor inversion symmetries, more complex band crossings can be pinned at high symmetry lines or points that feature larger Chern numbers than Weyl semimetals. For example, twofold crossings with quadratic or cubic dispersion are predicted to host |C|=2 or |C|=3 (*13, 14*). In materials with negligible spin-orbit coupling (SOC), three-fold and fourfold crossings can be found with |C|=2 per spin, whereas the combination of non-symmorphic symmetries and significant SOC gives rise to protected fourfold and sixfold degeneracies with Chern numbers up to a maximal magnitude of 4. The symmetry classification is exhaustive (*8, 15–18*) and predicts that |C| = 4 only occurs thanks to SOC and is the highest possible Chern number achievable for a multifold node in non-magnetic chiral topological semimetals. For linear band crossings in magnetic materials, the maximal Chern number is also 4 (*19*).

The family of chiral semimetals in space group 198, including RhSi, CoSi, AlPt, and PdBiSb, is expected to display these type of maximal |C| = 4 crossings, realized as a fourfold spin S=3/2 crossing at the Γ point (known as the Rarita–Schwinger fermion) and a sixfold S=1 crossing at the R point of the Brillouin zone, respectively. Despite several recent angle-resolved photoelectron spectroscopy (ARPES) experiments on all these candidates (*20–24*), the absolute magnitude of the Chern number, measured by counting the number of Fermi-arcs, has not yet been observed for two reasons. The first is that SOC in some of these materials is low and spin-split Fermi-arcs cannot be resolved, effectively leading to only two observable arcs. Most photoemission studies to date (*20–22*) in fact classify these nodes as having |C| = 2, whereas recent all-optical measurements (*25*) find a Chern number close to 4. The second reason is the difficulty in preparing clean and flat surfaces by cleaving or sputtering and annealing, which has resulted in rough or nonstoichiometric surfaces for all previously examined candidates, causing band broadening that can wash out signatures of spin-split bands caused by SOC. In this work, we overcome these obstacles by investigating a different chiral topological semimetal candidate PdGa from space group 198; this material has substantial SOC and can be prepared with flat, clean, and well-ordered surfaces by polishing and subsequent sputtering and annealing in ultrahigh vacuum (*26–28*). Employing ARPES and ab-initio calculations, we can clearly resolve the presence of multifold crossings in the bulk electronic structure of PdGa, as well as four topological Fermi-arcs on its surface, thus observing an experimental realization of the maximal Chern number |C| = 4. Interestingly, PdGa is known as an important catalyst, for instance for the semi-hydrogenation of acetylene (*29*), and shows potential for enantioselective catalytic reactions of chiral molecules (*30*). Because the Fermi-arcs are mostly derived from d-orbitals of Pd that are well-known to be important for catalysis (*31*), they enlarge the reservoir of catalytically active charge carriers at the sample surface where chemical reactions take place. Additionally, the topological protection of non-zero Chern numbers could suppress passivation of the Fermi-arcs, e.g. by hydrogenation (*32*).

The PdGa samples used in this study crystallize in the cubic space group 198 with a lattice constant of a=4.896 Å. The chiral motif in their structure is the helical arrangement of Pd and Ga atoms along the (111) direction (Fig. 1A). Upon a mirror operation, these helices reverse their handedness, which can be used to distinguish the two enantiomers of PdGa. We grew two enantiopure specimens of PdGa with opposite chirality via a self-flux method with a chiral seed crystal and used x-ray diffraction and the Flack method to determine the structural chirality of our samples, indicating almost ideal homochirality. More information about the refinements can be found in (*33*). The chirality of the crystal structure close to the surface can also be observed from the intensity distribution of low energy electron diffraction (LEED) patterns of the (100) surface (*28*) at an electron energy of $E_{kin}$=95 eV (Fig. 1B). As can be expected, the S-shaped intensity distribution is mirrored when comparing the two enantiomers. The crystals used for the ARPES and LEED studies were prepared by the same sputter-annealing recipe, which is well known to produce clean and stoichiometric surfaces of PdGa (*26*). In Fig. 1C, we display the results of an ab-initio bulk band structure calculation, which shows fourfold and sixfold band crossings at the Γ and R high symmetry points, respectively. Such band crossings in space group 198 were predicted to carry a Chern number of magnitude 4, with opposite signs at the Γ and R points (*15–18*). Because the Berry curvature is a pseudovector, a mirror operation will reverse the sign of the Chern numbers associated with the nodes at the high symmetry points. Such a mirror operation also leads to a reversal of the propagation direction of the Fermi-arcs (Fig. 1D). The multifold fermions at the Γ and R points act as sources (positive Chern number) or sinks (negative Chern number) of Berry curvature. One can imagine integrating the Berry flux passing through a two-dimensional slice that is dividing the Brillouin zone between the Γ and R points (blue shaded planes in Fig. 1D). Because of time-reversal symmetry, the Chern number of the slice is equivalent to half of the Chern number associated with the multifold fermions at Γ and R, and the sign of their Chern number depends on the direction of Berry flux. If we imagine this slice to be a two-dimensional quantum Hall phase, then the number of edge states of the slice is directly related to its Chern number magnitude, whereas their direction depends on its Chern number sign. The observation of a Fermi-arc doublet that is connecting the Γ and R points is, therefore, an unambiguous signature of a Chern number with magnitude 4, and the observation of the reversal of the Fermi-arc velocity is an unambiguous signature of a change in the Chern number sign associated with the multifold fermions.

We performed bulk sensitive soft X-ray ARPES measurements on the (100) surface of our PdGa samples to investigate their bulk electronic structure (Fig. 2). We find that multifold crossings predicted at the R- and Γ-points are indeed present (see Fig 2, A-C), and that our ab-initio calculations are in good qualitative agreement with the observed band dispersions. This agreement can also be observed from the Fermi surfaces for different high-symmetry planes displayed in Fig. 2, D-E. Further analysis of the spin-orbit splitting of the bulk bands can be found in (33).

After establishing the existence of multifold band crossings in PdGa, we will now investigate the topological character of these crossings via surface sensitive ARPES of the (100) surface of enantiomer A at low photon energies (hv<150 eV), as well as ab-initio slab calculations. By comparing the calculated and experimental Fermi surfaces in Fig. 3, A and B, we can identify the existence of Fermi arc surface states (indicated by red arrows). They thread through the projected bulk band gap (white areas indicated by blue lines in Fig 3A) and connect the projected bulk band pockets centered at $\bar{\Gamma}$ and $\bar{R}$. By performing photon energy dependent ARPES along the $\bar{R}$-$\bar{\Gamma}$-$\bar{R}$ direction, we confirm experimentally that these Fermi-arcs are indeed surface states without noticeable dispersion along the $k_z$ direction (perpendicular to the sample surface), as can be seen from Fig. 3C. Interestingly, we also find additional surface states that overlap with the projected bulk pocket at $\bar{\Gamma}$ (indicated by purple arrows). Owing to the sizable SOC in PdGa and high resolution of our ARPES data, we are furthermore able to resolve a spin-splitting in the surface Fermi-arcs (see Fig. 3, D-F, and the calculation in Fig. 3A for comparison). We can therefore conclude that four Fermi-arcs are connecting the projections of the multifold fermions located at $\bar{\Gamma}$ and $\bar{R}$ points. which constitutes an experimental confirmation of their maximal Chern number magnitude 4. We find that the SOC splitting of the Fermi-arcs close to the Fermi level is ~0.015 Å$^{-1}$ and ~60 meV. Because these multifold crossings are a generic feature of many chiral topological semimetals, we expect that our finding will also hold for other compounds from the same material family.

Next we investigate how the maximal Chern number in PdGa can be controlled by tuning the handedness of its crystal structure. When comparing the Fermi-surfaces for enantiomers A and B (Fig. 4A), we see that the Fermi arcs wind around the bulk pocket at $\bar{R}$ in opposite directions. By comparing the band dispersion of the Fermi-arcs between the two enantiomers along a line cut (Fig. 4C), we can see that the Fermi velocity of the edge states is indeed reversed, which implies that the Chern number signs are reversed between the two enantiomers (dispersions along a different direction can be found in (33)). This observation shows that the sign of the Chern numbers in topological semimetals can be controlled by deliberately choosing a sample with a specific handedness for experiments. We expect that this finding will serve as a control parameter in experiments that investigate the response of topological semimetals to external perturbations, such as all-optical measurement of the quantized circular photogalvanic effect (*25*). Here, a comparison of the nonlinear response between two enantiomers should give the same magnitude of the mesa-like plateau region in the photocurrent spectrum, albeit with a reversed sign. We furthermore expect eight counterpropagaing topological edge modes at a domain wall between enantiomers in PdGa, given that the Chern numbers for positive and negative momenta changes by 4 (33). The coupling of multifold fermions with opposite Chern number at this boundary could realize an interface Fermi-surface that is qualitatively different from the boundary to the vacuum and thereby enable distinct topological and correlated phenomena.

**Acknowledgments**

We would like to thank Leonard Nue, Andreas Pfister, and Lukas Rotach for excellent technical support. We acknowledge the Paul Scherrer Institut, Villigen, Switzerland for provision of synchrotron radiation beam time at beamline ADRESS of the SLS. We also acknowledge the Diamond Light Source for time on Beamline I05 under Proposal SI24703 and SI20617. N.B.M.S. would like to thank Adolfo G. Grushin for valuable discussions and Dieter Schöpplein for inspiration at early stages of the project. M.S. would like to thank Susann Scharsach for the DTA and DSC measurements.

**Funding:** K. M. and C. F. acknowledge financial support from the European Research Council (ERC) Advanced Grant No. 291472 "Idea Heusler" and 742068 "TOP-MAT"; and Deutsche Forschungsgemeinschaft (Project-ID 258499086 and FE 633/30-1). N.B.M.S. was supported by Microsoft. M.G.V. acknowledges support from DFG INCIEN2019-000356 from Gipuzkoako Foru Aldundia and the Spanish Ministerio de Ciencia e Innovacion (grant number PID2019-109905GB-C21). S.S. and R.W. acknowledge funding from the Swiss National Science Foundation under SNSF project number 159690. D.P. acknowledges the support from Chinese Scholarship Council. J.A.K. acknowledges support by the Swiss National Science Foundation (SNF-Grant No. 200021_165910).

**Author contributions:** N.B.M.S. conceived and coordinated the project, and performed the ARPES experiments and data analysis with support from S.S., J.A.K., D.P., and V.N.S. K.M. grew and characterized the samples with support from H.B. and M.S. M.G.V. performed the ab-initio calculations. F.J. and B.B. provided theoretical support. V.N.S., T.S., P.D., T.K.K., and C.C. maintained the ARPES endstations and provided experimental support. N.B.M.S. wrote the paper with support from F.J. and input from other co-authors. R. W., V.N.S., and C.F. supervised parts of the project.

**Competing interests:** The authors declare no competing interests.

**Data and materials availability:**

The data presented in this work is available on the PSI Public Data Repository (34).

**Supplementary Materials:**

Materials and Methods

Supplementary Text

Figs. S1-S4

Tables S1

References (35)-(50)

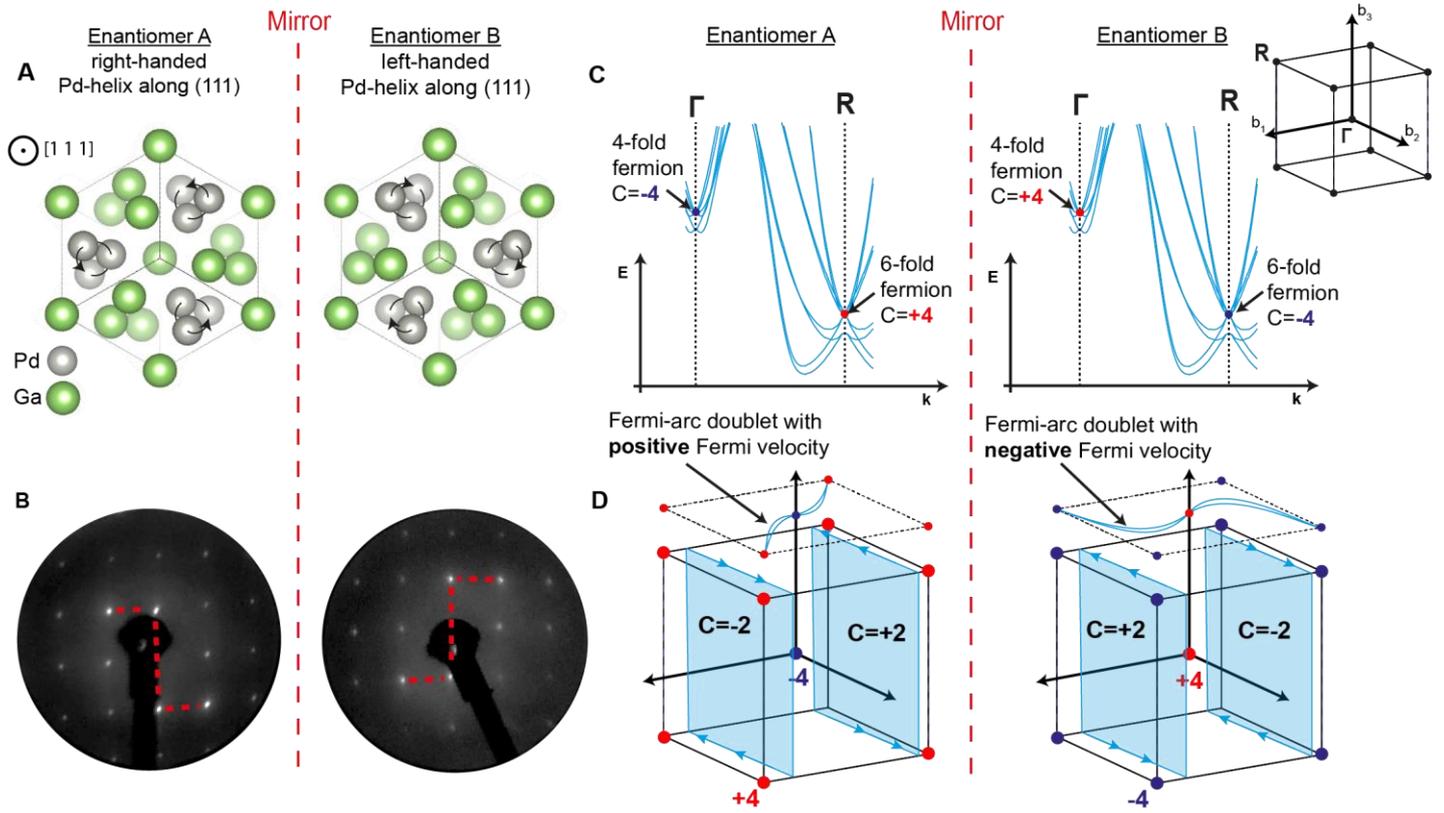

**Fig. 1: Structural and electronic chirality in the two enantiomers of PdGa**

- **(A)** Illustration of the crystal structure of two enantiomers of PdGa with opposite handedness.
- **(B)** Low energy electron diffraction patters for two samples with opposite chirality, measured with an electron energy of $E_{kin}$ = 95 eV. The S-shaped intensity distribution of the diffraction spots (highlighted by red dashed lines as guides for the eye) reflects the handedness of the crystal structure.
- **(C)** Ab-initio calculations of the band structure in PdGa, showing fourfold and sixfold band crossings at the Γ and R points. The Chern numbers associated with the crossings are of magnitude 4 and flip their sign upon a mirror operation. This reverses the direction of Berry flux that is flowing from the crossing with positive Chern number (red circles) towards the crossing with negative Chern number (blue circles). Inset shows the cubic Brillouin zone with high symmetry point Γ at the zone center and R at the zone corner.
- **(D)** Illustration of bulk boundary correspondence for PdGa and related chiral topological semimetals. Blue shaded slices indicate 2D quantum Hall phase with Chern numbers of magnitude 2. Dashed black lines indicate the edges of the surface Brillouin zone, and solid blue lin es and black arrows indicate the Fermi-arc surface states that are connecting the projections of R and Γ points.

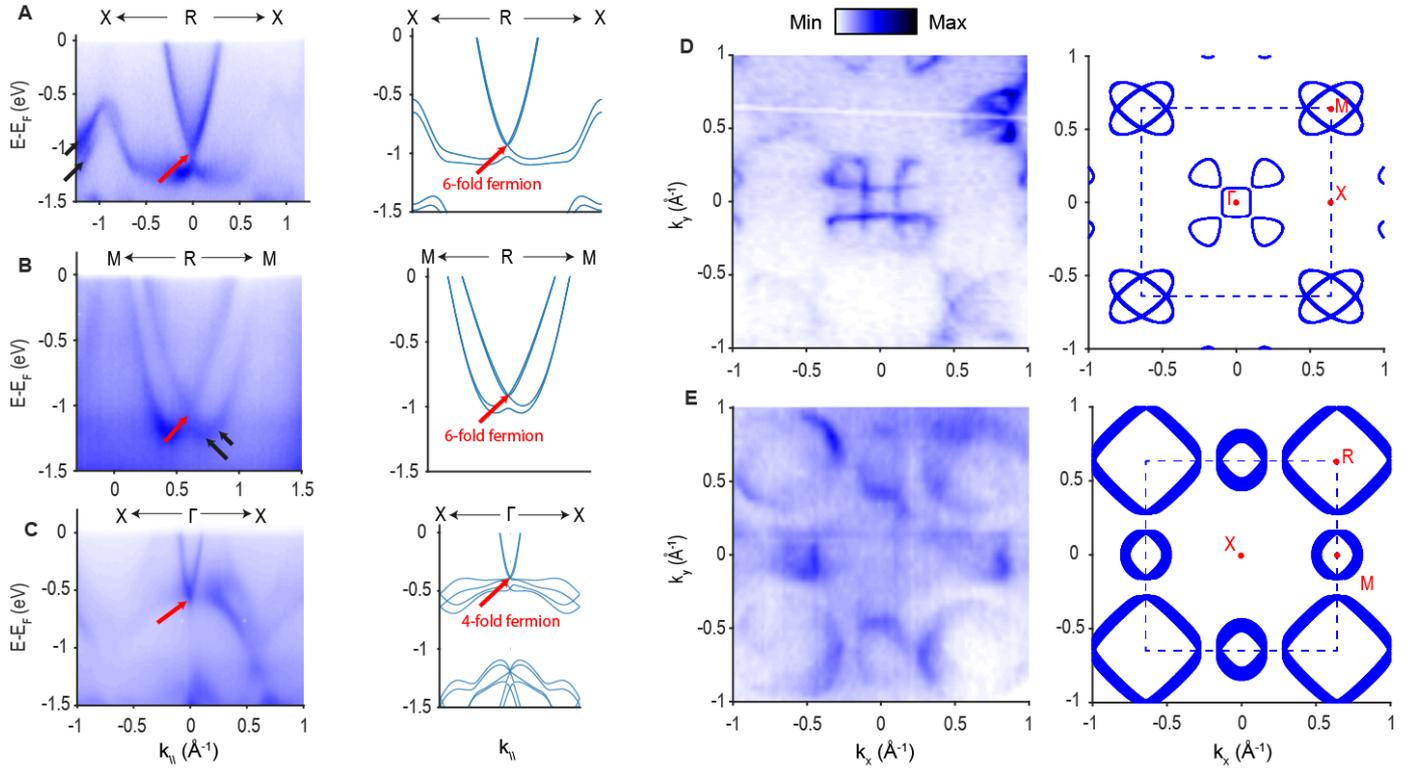

**Fig. 2: Electronic characterization of the bulk electronic structure of PdGa measured on the (100) surface of enantiomer A**

(A-C) Comparison between ARPES spectra (left) and ab-initio calculations (right) along high symmetry lines that pass through the Γ and R points. Multifold fermions are indicated by red arrows and spin-orbit splitting is indicated by black arrows. Spectra were measured at hv=550 eV, 552 eV, and 620 eV, respectively, all with linear-vertical polarization (LV).

(D-E) Comparison between experimental Fermi-surfaces (left) and ab-initio calculations (right). (D) shows the Fermi surface in the high symmetry plane containing the G point (measured with hv=620 eV, LV polarization), whereas (E) contains the plane including the R point (measured with hv=540 eV, LV polarization). The blue dashed lines indicate the boundary of the bulk Brillouin zone. Ab initio calculations include $k_z$ broadening of 0.1 Å$^{-1}$.

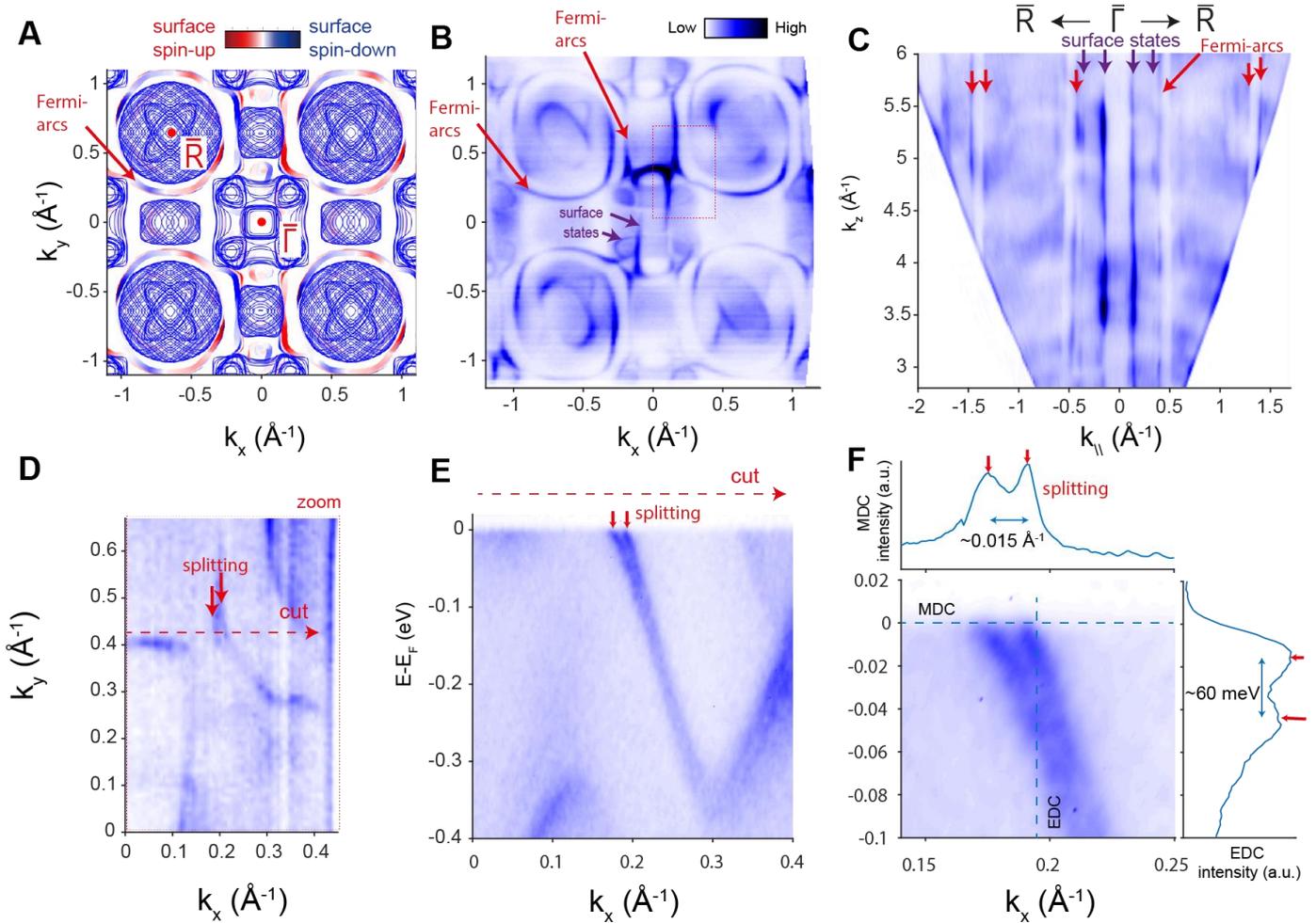

**Fig. 3: Surface electronic structure of the (100) surface of enantiomer A**

**(A)** Ab-initio slab calculation of the Fermi surface in the (100) plane (i.e. $k_x$ vs. $k_y$ plane) showing surface Fermi-arcs (indicated by a red arrows), superimposed by projected bulk band structure calculation (solid blue lines).

**(B)** Experimental Fermi surface measured with hv=60 eV and linear-horizontal (LH) polarization. Red arrows indicate Fermi-arcs, whereas purple arrows indicate additional surface states that overlap with the projected bulk states at $\bar{\Gamma}$.

**(C)** Experimental Fermi surface perpendicular to the sample surface (i.e. $k_y$ vs. $k_z$ plane), showing that the Fermi-arcs and surface states (indicated by red and purple arrows, respectively) show negligible dispersion along the $k_z$ direction. Conversion from photon energy was performed within free-electron final state approximation with inner potential of $V_0 = 12$ eV.

**(D)** Magnified Fermi-surface measured in the region of the red dotted rectangle shown in (B) with hv= 30 eV and LH polarization. Red arrows indicate spin-splitting of Fermi-arcs

**(E)** Band dispersion measured along the path in momentum space indicated by the dashed red arrow shown in (D) that is crossing the Fermi-arcs. Red arrows indicate their spin-splitting.

**(F)** Magnified version of (E), insets are momentum distribution curve (MDC) and energy distribution curve (EDC) along the dashed blue lines

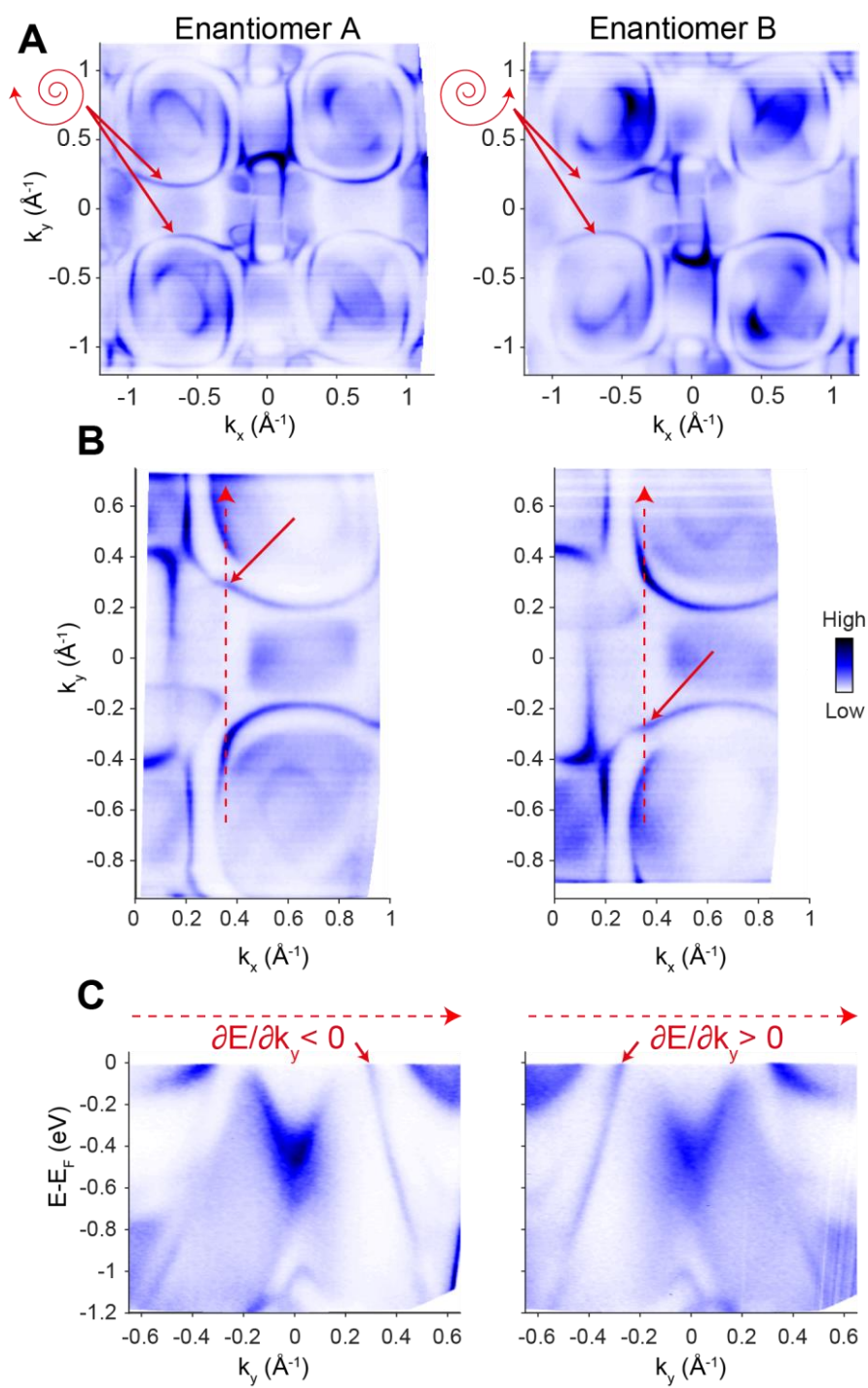

**Fig. 4: Comparison of the surface electronic structure of the (100) surface of enantiomer A (left) and enantiomer B (right)**

(A) Comparison of the Fermi-surfaces for two enantiomers, measured with photon energy hv=60 eV and LH polarization. Red arrows indicate Fermi-arcs that reverse the direction along which they are dispersing around the $\bar{R}$ pocket under a mirror operation.

(B) Comparison of magnified Fermi-surfaces measured with photon energy hv=30 eV and LH polarization. Red dashed line indicates momentum path shown in (C). Red solid arrows indicate

Fermi-arcs that are crossing the projected bulk band gap that separates the projected bulk pockets at $\bar{\Gamma}$ and $\bar{R}$.

**(C)** Band dispersion along the path indicated by the red dashed line in (B). Red arrows indicate the Fermi-arcs that are crossing the projected bulk band gap. One can see that the component of the fermi velocity along the $k_y$ direction switches sign between the two enantiomers.